\documentclass[preprint,3p,twocolumn,sort&compress]{elsarticle}

\usepackage{lineno,hyperref}
\modulolinenumbers[1]
\usepackage{color}
\usepackage{amsmath}
\usepackage{amssymb}
\usepackage{graphicx}
\journal{Optics Communications}

\begin{document}

\begin{frontmatter}

\title{Tunable Four-Wave Mixing in AlGaAs Waveguides of Three Different Geometries}


\author[1]{Daniel H. G. Espinosa}

\author[1]{Kashif M. Awan\fnref{3}}
\fntext[3]{Present address: Stewart Blusson Quantum Matter Institute, University of British Columbia, V6T 1Z4, Vancouver, Canada.}

\author[1]{Mfon Odungide}

\author[2]{Stephen R. Harrigan\fnref{4}}
\fntext[4]{Present address: Institute for Quantum Computing and Department of Physics \& Astronomy, University of Waterloo, Waterloo, Ontario N2L 3G1, Canada.}

\author[1]{David R. Sanchez J.\fnref{5}}
\fntext[5]{Present address: Karlsruhe School of Optics \& Photonics, Karlsruhe Institute of Technology, Kaiserstrasse 12, Karlsruhe, 76131, Germany.}

\author[1,2]{Ksenia Dolgaleva\corref{mycorrespondingauthor}}
\cortext[mycorrespondingauthor]{Corresponding author}
\ead{ksenia.dolgaleva@uottawa.ca}

\address[1]{School of Electrical Engineering and Computer Science, University of Ottawa, Advanced Research Complex, 25 Templeton St., Ottawa, Ontario, K1N 6N5, Canada.}

\address[2]{Department of Physics, University of Ottawa, Advanced Research Complex, 25 Templeton St., Ottawa, Ontario, K1N 6N5, Canada.}

\begin{abstract}
The AlGaAs material platform has been intensively used to develop nonlinear photonic devices on-a-chip, thanks to its superior nonlinear optical properties.
We propose a new AlGaAs waveguide geometry, called half-core etched, which represents a compromise between two previously studied geometries, namely the nanowire and strip-loaded waveguides, combining their best qualities.
We performed tunable four-wave mixing (FWM) experiments in all three of these geometries in the telecommunications C-band (wavelengths around $1550~\text{nm}$), with a pulsed pump beam and a continuous-wave (CW) signal beam. The maximum FWM peak efficiencies achieved in the nanowire, strip-loaded and half-core geometries were about $-5~\text{dB}$, $-8~\text{dB}$ and $-9~\text{dB}$, respectively. These values are among the highest reported in AlGaAs waveguides. The signal-to-idler conversion ranges were also remarkable: $161~\text{nm}$ for the strip-loaded and half-core waveguides and $152~\text{nm}$ for the nanowire. Based on our findings, we conclude that the half-core geometry is an alternative approach to the nanowire geometry, which has been earlier deemed the most efficient geometry, to perform wavelength conversion in the spectral region above the half-bandgap.
Moreover, we show that the half-core geometry exhibits fewer issues associated with multiphoton absorption than the nanowire geometry.
\end{abstract}


\end{frontmatter}


\section{Introduction}
Optical communication networks nowadays rely on both optics and electronics, where the signal processing is performed mainly in the electrical domain, while the long-distance data transmission along with the switching and multiplexing of the wavelength channels occur in the optical domain. Intense research and development of photonic integrated circuits have been conducted to eliminate many optical-to-electrical and back-to-optical (OEO) conversions in optical communication networks by enabling signal processing in the optical domain, \textit{i.e.}, all-optical signal processing~\cite{yan2012all,eggleton2012photonic,rusing2019toward,ji2019all}.
The idea of all-optical signal processing relies on integrated optical devices with strong nonlinear optical interactions, in contrast to the conventional integrated technologies for the transmitters and receivers that do not require optical nonlinearity~\cite{ji2019all}. Among many material platforms under development for all-optical signal processing, the III-V semiconductor aluminum gallium arsenide (AlGaAs) stands out due to its superior nonlinear optical properties~\cite{stegeman1994algaas,aitchison1997nonlinear}.
It is transparent in the communication C-band, which minimizes the linear absorption loss. Moreover, it has a direct bandgap, which is useful for the monolithic integration of active and passive devices. Further, its optical properties are tunable by choice of the aluminum content or fraction, $x$, in the Al$_x$Ga$_{1-x}$As compound~\cite{gehrsitz2000refractive,adachi1988optical}.

There are many reports on aluminum gallium arsenide waveguides based on a three-layer arrangement of Al$_x$Ga$_{1-x}$As with different compositions epitaxially grown on a gallium arsenide (GaAs) substrate~\cite{johnson2019orthogonal,kultavewuti2017polarization,kultavewuti2016correlated,Dolgaleva2015,kultavewuti2015low,porkolab2014low,lacava2014nonlinear,wathen2014efficient,apiratikul2014enhanced,mahmood2013polarization,cannon2013all,wang2012ultra,dolgaleva2011compact,dolgaleva2010broadband}, which is one of the earliest AlGaAs waveguide technologies~\cite{shelton1979characteristics,blum1974optical}.
The different aluminum content of each layer provides the refractive index contrast necessary for the light confinement. Alternative AlGaAs technologies have been developed in recent years, such as AlGaAs/Al$_\text{x}$O$_\text{y}$~\cite{li2011second,scaccabarozzi2006enhanced},  AlGaAs-on-AlO$_\text{x}$~\cite{carletti2017controlling,ozanam2014toward}, AlGaAsOI (AlGaAs-on-insulator)~\cite{pu2018ultra,da2017characterization,ottaviano2016low,pu2016efficient}, AlGaAsOS (AlGaAs-on-sapphire)~\cite{zheng2018high},  GaAs/AlGaAs~\cite{haas2019mid,maltese2018towards,sieger2017portable,haas2016sensing,sieger2016optimizing,wang2014gallium,apiratikul201010}, suspended GaAs~\cite{stievater2014mid} and suspended AlGaAs on GaAs~\cite{roland2020second} and on silicon substrates~\cite{chiles2019multifunctional}. Enhanced second-harmonic generation~\cite{li2011second,scaccabarozzi2006enhanced} and optical parametric oscillation~\cite{ozanam2014toward} have been demonstrated in AlGaAs-on-Al$_\text{x}$O$_\text{y}$ technology. AlGaAsOI was inspired by silicon-on-insulator (SOI) technology, both in concept and name~\cite{pu2018ultra}. Examples of AlGaAsOI devices are high-quality and low-loss waveguides and microring resonators made of AlGaAs layer on a SiO$_2$ layer, defined on top of a GaAs substrate~\cite{ottaviano2016low}. The targeted applications of AlGaAsOI are frequency comb generation~\cite{pu2016efficient}, continuous-wave four-wave mixing (CW-FWM) for high-order quadrature amplitude modulation~\cite{da2017characterization} and broadband wavelength conversion~\cite{pu2018ultra}. AlGaAsOS relies on AlGaAs defined on an Al$_2$O$_3$ substrate, rather than on top of GaAs, through a wafer bonding process. It has the potential of operating in the mid-infrared (MIR) range~\cite{zheng2018high}. High-quality microring resonators and highly efficient CW-FWM have been demonstrated in this platform~\cite{zheng2018high}. For mid-infrared (MIR) operation, incidentally, there are other reported approaches: GaAs/AlGaAs thin-films slab~\cite{sieger2016optimizing,charlton2006fabrication} and ridge~\cite{wang2014gallium} waveguides and microring resonators~\cite{haas2019mid}. These devices are made of GaAs thin films deposited on an AlGaAs layer over a GaAs substrate. The targeted applications of GaAs/AlGaAs waveguide are infrared spectroscopy~\cite{sieger2017portable}, sensing~\cite{charlton2006fabrication,haas2016sensing,wang2012ultra}, quantum photonics~\cite{wang2014gallium} and controlled $\frac{\pi}{2}$-phase delay between the TE and TM modes that results in the generation of a circularly polarised mode~\cite{maltese2018towards}. In the suspended GaAs, AlGaAs-on-GaAs and AlGaAs-on-silicon technologies, the waveguide is suspended in an air layer, which increases the refractive index contrast. Second-harmonic, supercontinuum and difference-frequency generations in mid-IR were reported in such devices~\cite{roland2020second,chiles2019multifunctional,stievater2014mid}.

The newer AlGaAs technologies, such as AlGaAsOI and AlGaAsOS look very attractive due to their similarity to SOI, low propagation loss and high mode confinement. On the other hand, the conventional AlGaAs approach (Al$_x$Ga$_{1-x}$As layers on GaAs substrate) still represents practical interest because commercial III-V foundries support this technology. Moreover, the fabrication processes needed for newer technologies are more complex and would largely complicate integrated optical circuit manufacturing. This makes the conventional AlGaAs technology appealing to the applications in optical communication networks: the existing commercial technology for integrated optical chips would require insignificant modifications to incorporate optical signal processing devices. Besides, the conventional AlGaAs technology has been proven to work well for integrated sources of quantum light, suitable for the applications in secure communications and quantum computing~\cite{kultavewuti2017polarization,autebert2016integrated,kultavewuti2016correlated,kang2015two,sarrafi2014high,sarrafi2013continuous,orieux2013direct}.

All-optical signal processing can be realized through the four-wave mixing (FWM) process whereby two input photons with the frequencies $\omega_{1}$ and $\omega_{2}$ are destroyed, and two output photons with the frequencies $\omega_{3}$ and $\omega_{4}$ are created. Since the total energy must be conserved, the frequencies of the photons are related as $\omega_{1} + \omega_{2} = \omega_{3} + \omega_{4}$. In the case of degenerate four-wave mixing (DFMW), two pump photons with the frequency $\omega_{\text{p}}$ are converted into one signal photon of frequency $\omega_{\text{s}}$ and one idler photon of frequency $\omega_{\text{i}}$, related by $2 \omega_{\text{p}} = \omega_{\text{s}} + \omega_{\text{i}}$~\cite{agrawal2000nonlinear}. DFWM has been studied in waveguides and microring resonators of different geometries and material platforms, such as AlGaAs, silicon, and chalcogenide glasses. Comparisons between different material platforms for DFWM can be found in Refs.~\cite{pu2018ultra,yang2018high,dolgaleva2011compact}. It is evident from the detailed comparison conducted in these sources that AlGaAs exhibits superior linear and nonlinear optical properties and holds promise for nonlinear integrated photonic devices. This serves as a motivation for many researchers to continue investigating this material platform and its further development for all-optical signal processing and other applications.

\begin{figure*}[htbp]
\centerline{\includegraphics[width=17cm]{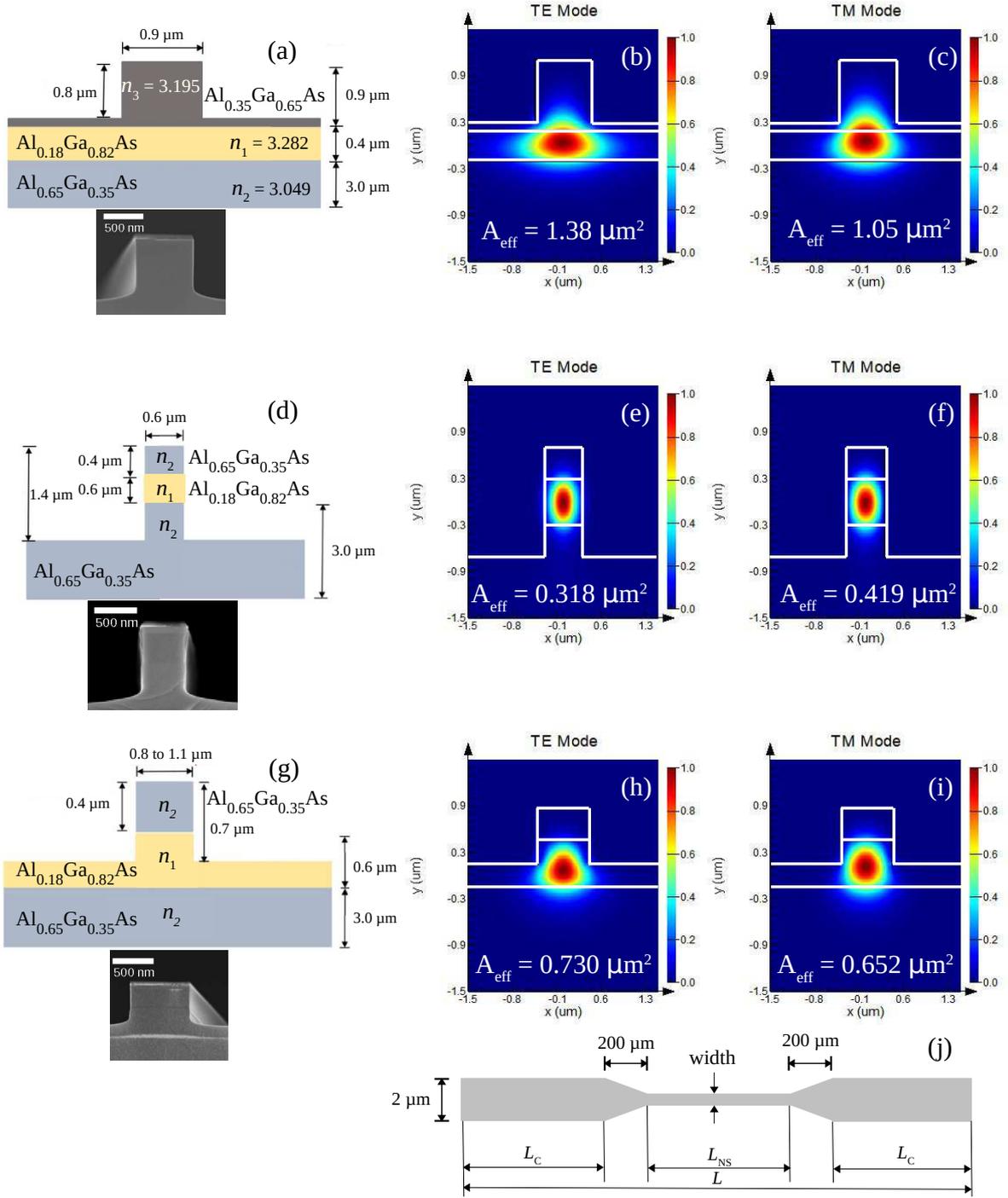}}
\caption{\label{fig:samples}(color online) Cross-sectional geometries and material compositions of (a) strip-loaded, (d) nanowire and (g) half-core waveguides, respectively. The refractive index at the wavelength $1500~\text{nm}$ is indicated for each layer by $n_1$, $n_2$ or $n_3$~\cite{gehrsitz2000refractive}. The core layers are represented in beige color; the insets show SEM images of the corresponding waveguide cross-sections. The intensity distributions of the fundamental TE and TM modes for $\lambda = 1500~\text{nm}$ are presented in (b) and (c), respectively, for the strip-loaded, in (e) and (f) for the nanowire and in (h) and (i) for the half-core geometries. Part (j) shows the top-view of all our devices. The width varies from $0.6$ to $1.1~\mu\text{m}$, and the length $L_{\text{NS}}$ varies from $0.5$ to $4~\text{mm}$. The total length of the sample $L$ was $5.26~\text{mm}$, $5.33~\text{mm}$, and $5.87~\text{mm}$ for the strip-loaded, nanowire and half-core devices, respectively. The length of the coupling part is calculated by $L_\text{C} = (L - L_\text{NS} - 0.4~\text{mm})/2$.}
\end{figure*}

In this work, we realize a new AlGaAs waveguide geometry, the half-core-etched waveguide, and report the comparative experimental study of FWM in the new and two previously studied AlGaAs waveguide geometries. The first AlGaAs waveguide geometry, studied earlier, is called the strip-loaded, ridge, or rib~\cite{dolgaleva2010broadband,dolgaleva2011compact}. It consists of a planar slab waveguide comprising a high-index core layer sandwiched between low-index upper and lower claddings and a ridge (strip) defined on top of the upper cladding layer. The ridge and the upper cladding are made of the same material, and the light is confined and propagates in the core area just below the ridge [see Fig.~\ref{fig:samples} (a), (b), and (c)]. In another variation of the ridge geometry, the upper cladding and the core layers are completely etched, except for the region below the ridge. The CW-FWM, performed with CW pump and signal beams, is presented in Ref.~\cite{apiratikul201010} for this type of waveguides. The second geometry is the pedestal or vertical waveguide, where the core layer is completely etched through, and so is part of the lower cladding layer. The guided mode resides in the pedestal, and the high refractive index contrast between the core's sidewalls and the air allows for a smaller effective mode area [see Fig.~\ref{fig:samples} (d), (e), and (f)]. This geometry is also called nanowire when the width of the waveguide (pedestal) is sub-$\mu$m. DFWM in a pedestal geometry was first demonstrated in microring resonators~\cite{absil2000wavelength,van2002optical}. Later, FWM was demonstrated in $1.2-\mu$m-wide waveguides~\cite{cannon2013all,wathen2014efficient} and in nanowire waveguides~\cite{dolgaleva2013highly,Dolgaleva2015}. CW-FWM in nanowire waveguides~\cite{lacava2014nonlinear,wathen2014efficient,apiratikul2014enhanced} and microring resonators~\cite{kultavewuti2015low} has also been reported. Although the nanowire geometry exhibits higher propagation loss than the strip-loaded waveguide, the light is confined to a smaller area, leading to higher optical nonlinearities in a nanowire waveguide for the same level of the input power. Furthermore, propagation loss can be reduced with improvements in the fabrication process that would allow one to minimize sidewall roughness~\cite{porkolab2014low}.

In another study, dispersion management, which is possible in nanowire waveguides~\cite{meier2007group}, in combination with the natural birefringence of AlGaAs, was used to demonstrate polarization-insensitive FWM in carefully designed nanowire waveguides~\cite{cannon2013all,mahmood2013polarization}. Moreover, FWM with an orthogonally polarized pump and signal (orthogonal FWM) has been proposed, enabling the removal of the output pump beam without affecting the signal and idler beams. The results of the simulations indicate that an orthogonal FWM can be implemented in nanowire waveguides designed to have the same effective index values for the transverse electric (TE) and transverse magnetic (TM) propagation modes~\cite{johnson2019orthogonal}.

A compromise between the strip-loaded and nanowire geometries would combine the good confinement of the nanowire with the low propagation loss of the strip-loaded. For that purpose, the half-core geometry concept was first presented in~\cite{awan2015aluminium}, with some simulations of the propagating modes, but no experimental results have been reported to date. Following that effort, in this work, we report the details of the design, simulations, fabrication, and characterization of AlGaAs half-core waveguides, and demonstrate, for the first time, FWM in this geometry. In addition, we present the details of the fabrication, optical characterization and FWM results for the nanowire and strip-loaded waveguides fabricated from the same wafer, which allows for comparing the FWM efficiencies of these three geometries.

\section{AlGaAs Waveguides}

\subsection{Design}

GaAs was selected as the substrate material for the growth of the Al$_{x}$Ga$_{1-x}$As layers because there is excellent lattice matching between GaAs and all possible AlGaAs compositions~\cite{bett1999iii,wasilewski1997composition}. For instance, the lattice mismatch between Al$_{0.65}$Ga$_{0.35}$As and GaAs is expected to be $9.75\nobreak\times \nobreak10^{-4}$~\cite{wasilewski1997composition}. The waveguide core composition was chosen to be Al$_{0.18}$Ga$_{0.82}$As so that the two-photon absorption (2PA) around the Telecommunication C-band wavelength $1550~\text{nm}$ is negligible. 2PA is only significant for the photon energies higher than half-the-bandgap, which corresponds to $\lambda \lessapprox 1500~\text{nm}$ for Al$_{0.18}$Ga$_{0.82}$As~\cite{bosio1988direct,oelgart1987photoluminescence,lee1980electron,casey1978room}, and this particular composition has already been used in previous studies in order to avoid 2PA~\cite{stegeman1994algaas}. The compositions of the upper and lower cladding layers should have the highest possible refractive index contrast with the core layer. Since the refractive index decreases with the aluminum concentration ($x$)~\cite{gehrsitz2000refractive}, the maximum possible $x$ should be used. As it is known that the oxidation of AlGaAs compounds occurs when the aluminum concentration is $x > 0.7$, $x = 0.65$ was chosen to maximize the refractive index contrast while avoiding the oxidation.

The waveguide cross-sectional dimensions and aluminum content of the cladding layers were then optimized through a series of simulations performed with the finite difference eigenmode (FDE) solver of $\textit{Lumerical Mode Solutions}$, using the dispersion characteristics of Al$_{x}$Ga$_{1-x}$As given in~\cite{gehrsitz2000refractive}. The lower cladding thickness was set to $3~\mu\text{m}$ to avoid the mode leaking into the higher-index GaAs substrate. The design was considered optimal when the effective mode area ($A_\text{eff}$) was minimized while the propagation mode remained mainly in the core layer. The expression used to calculate $A_\text{eff}$ in FDE solver is based on the electric field amplitude distribution within the waveguide's cross-section in the $xy$-plane, $E(x,y)$, and is given by~\cite{agrawal2000nonlinear}
\begin{align}
A_\text{eff} & = \frac{ \left[ \int_{-\infty}^\infty \int_{-\infty}^\infty |E(x,y)|^2 \text{d}x \text{d}y \right]^2}{ \int_{-\infty}^\infty \int_{-\infty}^\infty |E(x,y)|^4\text{d}x \text{d}y }.\label{eq:aeff}
\end{align}

Eq.~(\ref{eq:aeff}) assumes that the refractive index does not change significantly within the waveguide's cross-section, which is an approximation. This equation was adopted in this work for the practical convenience of numerical calculations with the FDE solver. A more precise expression for the effective mode area should include the refractive indices of different layers as weighting factors~\cite{yang2018invited}.

The simulations were performed by varying the following parameters for each geometry, one at a time: the core thickness, the upper cladding thickness, the upper and lower cladding aluminum content and the etching depth. The optimized dimensions and layer compositions with the lowest $A_\text{eff}$ are presented in Fig.~\ref{fig:samples}~(a), (d) and (g) for the strip-loaded, nanowire, and half-core geometries, respectively. Note that for the strip-loaded geometry, the best aluminum concentration in the upper cladding layer is $x = 0.35$, which corresponds to a lower refractive index in this layer than in the core. More details of the optimization study can be found in~\cite{awan2018fabrication}.

The simulated electric field distributions and values of $A_\text{eff}$ are also presented in Fig.~\ref{fig:samples}. In the strip-loaded geometry, the light propagates below the ridge (load) while the core layer (strip) spreads similarly to the core of a planar waveguide. The propagation mode is weakly confined in the core region just below the ridge [see Figs.~\ref{fig:samples}~(a), (b), and (c)]. In the nanowire geometry, the propagation mode is mostly confined to the vertical ridge (wire), where the bottom and top of the core are in contact with the cladding layers while the side walls are exposed to the air [see Figs.~\ref{fig:samples}~(d), (e), and (f)]. For the half-core geometry, the strip-loaded-like planar core layer is partially etched (``half-etched core''), and the mode is more confined than that of the strip-loaded geometry, but less than that of the nanowire [see Figs.~\ref{fig:samples}~(g), (h) and (i)].

To minimize the insertion loss, the input and output of the waveguides were made $2$-$\mu\text{m}$-wide for all the geometries, and the light was then coupled into the narrower sections by $200$-$\mu\text{m}$-long tapers, as shown in Fig.~\ref{fig:samples} (j). The length of each $2$-$\mu\text{m}$-wide coupling waveguide, or input and output coupler, is symbolized by $L_{\text{C}}$. $L_\mathrm{NS}$ and $L$, respectively, symbolize the length of the sub-$\mu\text{m}$-wide waveguide part and the total length. We call the part of the waveguide where both the width and height have sub-$\mu\text{m}$ dimensions ``nanosectional'' (NS). We also fabricated straight $2$-$\mu\text{m}$-wide waveguides of the same length $L$  on the chips with the devices during the same process, which we named reference waveguides, and used them to characterize the $2$-$\mu\text{m}$-wide parts of the devices.

\subsection{Fabrication process}

The substrate for all the devices was a $76$-mm-diameter, single-side polished GaAs (100) wafer with a miscut of $2^\text{o}$ towards the $110$-direction. 
The layers shown in Fig.~\ref{fig:samples} were grown on the substrate using molecular beam epitaxy provided by CMC Microsystems in a vacuum generator V80 system. 
A $0.1$-$\mu \text{m}$-thick GaAs capping layer was grown on top of the layer stack to prevent oxidation. Then the wafer was cleaved into $15~\text{mm}$ x $15~\text{mm}$ chips.

The next steps were electron beam (e-beam) lithography and etching. A $200$-nm-thick layer of silicon dioxide (SiO$_2$) was deposited by plasma-enhanced chemical vapour deposition over the chips, followed by a $50$-nm-thick chromium layer deposition by sputtering. Then, a $350$-nm-thick layer of negative-tone e-beam resist (maN-2403) was deposited by spin-coating at $4000~\text{rpm}$ for $1~\text{min}$ and pre-baked at $90~^\text{o}\text{C}$ for $1\:\text{min}$. The waveguides were then patterned on the resist using a $100$-kV e-beam lithography system operating with a dose of $210~\mu\text{C}/\text{cm}^2$, and the resist was developed for $40~\text{s}$ in the Ma-D 525 developer (from Micro Resist Technology GmbH). 

The waveguides patterned with the maN-2403 mask were then transferred to the chromium layer, SiO$_2$ layer and AlGaAs layers by successive inductively coupled plasma - reactive ion etchings (ICP-RIE). The following parameters were used for the etching: to etch chromium, $10~\text{sccm}$ of oxygen and $20~\text{sccm}$ of chlorine at $10~\text{mTorr}$, with RIE and ICP powers of $120~\text{W}$ and $500~\text{W}$, respectively; to etch SiO$_2$, $10~\text{sccm}$ of O$_2$ and $20~\text{sccm}$ of C$_4$F$_8$ at $6~\text{mTorr}$, with RIE and ICP powers of $300~\text{W}$ and $1000~\text{W}$, respectively; to etch Al$_x$Ga$_{1-x}$As, $15~\text{sccm}$ of BCl$_3$ and $5~\text{sccm}$ of Ar~\cite{maeda1999inductively,tamura1984gaas} at $15~\text{mTorr}$, with RIE and ICP powers of $75~\text{W}$ and $200~\text{W}$, respectively. The pressure and RIE and ICP powers values were determined after the optimization process.

Finally, we cleaved the chips with a diamond-edge cleaver to expose the waveguides' input and output facets. A cross-sectional scanning electron microscopy images of the facets are shown in the insets of Figs.~\ref{fig:samples}~(a), (d), and (g).

\section{Experimental}

\subsection{Four-wave mixing}

\begin{figure}[htbp]
\centerline{\includegraphics[width=9cm]{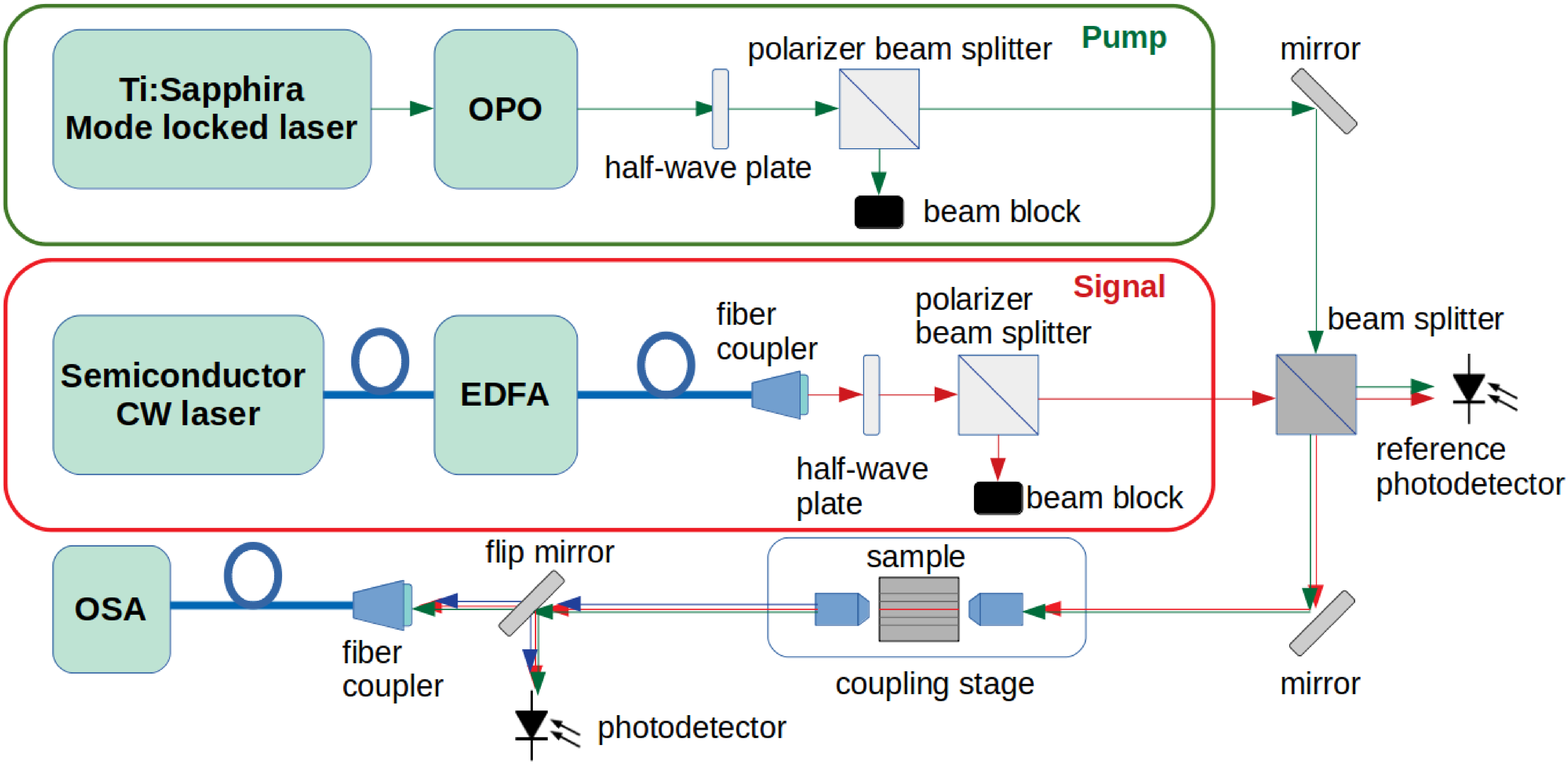}} \caption{\label{fig:diagram}(color on-line) Schematic of the experimental setup used for the degenerate four-wave mixing and nonlinear transmission experiments. OPO – optical parametric oscillator; CW – continuous-wave; EDFA – erbium-doped fiber amplifier; OSA – optical spectrum analyzer.}
\end{figure}

A schematic of the setup used to perform four-wave mixing experiments is presented in Fig.~\ref{fig:diagram}. The pump beam is the output of an optical parametric oscillator (OPO) (model Mira-OPO, from Coherent inc.) excited by a titanium sapphire laser (Mira-900, from Coherent inc.). The OPO repetition rate and pulse width is $f_\text{p} = 76.6~\text{MHz}$ and $\tau = 3~\text{ps}$, respectively. The signal beam is the output of a tunable semiconductor CW laser (model TSL-710, Santec inc.) amplified by an erbium-doped fibre amplifier (model AEDFA-33-B-FA, Amonics inc.). Polarizer beam splitters and half-wave plates controlled the output power and polarization of each beam. The signal and pump beams were coupled into and out of the waveguides using objective lenses and micropositioning stages. The light from the output was coupled into an optical spectrum analyzer (OSA, model AQ6315E, Ando inc.) using an optical fibre coupler. The coupling loss from the waveguide's output into the OSA was estimated at $6(1)~\text{dB}$, and it was accounted for in the calculation of FWM efficiency. The insertion (coupling) loss was estimated to be $5(1)~\text{dB}$, $12(1)~\text{dB}$, and $7(1)~\text{dB}$ for the strip-loaded, nanowire, and half-core devices, respectively. The incident signal and pump powers before coupling into the waveguide were approximately $28~\text{mW}$ ($14.5~\text{dBm}$) and $46~\text{mW}$, respectively.

\subsection{Linear loss measurement and nonlinear transmission technique}

The method of Fabry-Perot fringes was used to measure the linear propagation loss coefficient ($\alpha_1$) for the $2$-$\mu$m-wide references waveguides, as described in detail in~\cite{saeidi2018demonstration,Dolgaleva2015,tittelbach1993comparison}. In short, we performed a wavelength scan of a tunable CW semiconductor laser (model TSL-710, Santec inc.) in steps of $2~\text{pm}$ while measuring the transmitted power by photodiodes. The laser was coupled into and out of the waveguide using the objective lenses and micropositioning stages. The waveguide's input and output facets partially reflect the light so that the waveguide works as a Fabry-Perot cavity. The positions of the constructive and destructive interference maxima and minima depend on the wavelength, whereas the amplitudes depend on the waveguide length and linear loss.

To measure the linear loss coefficient of the nanosectional part ($\alpha_1^{\text{NS}}$), and to study the nonlinear loss, we used the nonlinear transmission technique with the setup similar to that used for the FWM experiments, but with the pulsed (pump) arm only. Rotating the half-wave plate, we changed the incident power ($P_\text{in}$) between $0$ and $45~\text{mW}$, while measuring the output power ($P_\text{out}$) with a photodiode and a power meter. $P_\text{out}$ subjects to both linear and nonlinear losses. We represent by $P_\text{out}^{\star}$ the output power in a hypothetical case where only the linear loss is present at the devices.

\subsubsection{Linear regime}

With low incident power ($0 < P_\text{in} \lessapprox 3~\text{mW}$), the sample demonstrated the presence of the linear loss only, and $P_\text{out} = P_\text{out}^{\star}$. Hence, the only loss sources for this power level were the linear propagation loss, characterized by the linear loss coefficient $\alpha_1$, and the coupling loss, represented by the coupling efficiency $\eta_{\text{C}}$. We used $\alpha_1$ and the reflectivity $R$ values, obtained from the Fabry-Perot method, as constants, and fitted the equation
\begin{align}
P_{\text{out}}^{\star} & = (1-R)^2 \eta_{\text{C}} e^{-\alpha_1 L} P_{\text{in}}\label{eq:PinPout}
\end{align}
to the $P_\text{out}$ vs.\ $P_\text{in}$ data collected from a $2$-$\mu$m-wide references waveguides to determine $\eta_{\text{C}}$. In Eq.~(\ref{eq:PinPout}), $R$ is the Fresnel reflection coefficient and $L$ is the length of the waveguide.

For the devices with a set of widths, as presented in Fig.~\ref{fig:samples} (j), we determined $\alpha_1^{\text{NS}}$ by fitting the equation 
\begin{align}
P_{\text{out}}^{\star} & = (1-R)^2 t_{\text{l}} \eta_{\text{C}}  e^{(-2 \alpha_1 L_{\text{C}} - \alpha_1^{\text{NS}} L_{\text{NS}})} P_{\text{in}} \label{eq:PinPoutNS}
\end{align}
to the $P_\text{out}$ vs.\ $P_\text{in}$ data, collected in the linear regime (at low values of incident power). As the input and output coupling parts of these waveguides have the same geometry as that of the reference waveguides, we used the same value of $\eta_{\text{C}}$, $\alpha_1$ and $R$ that we determined for the reference waveguides, and found $\alpha_1^{\text{NS}}$ from the fitting result.
In Eq.~(\ref{eq:PinPoutNS}), $t_{\text{l}}$ is the taper loss coefficient ($t_\text{l} \approx 0.8$ for the nanowire and $t_\text{l} \approx 1$ for the strip-loaded and half-core waveguides), $L_{\text{C}}$ is the total length of the $2$-$\mu$m-wide parts, and $L_{\text{NS}}$  is the length of the nanosectional part, respectively. The taper loss coefficient was estimated by comparing the output power of $2$-$\mu$m-wide reference waveguides and the output power of taper-to-taper ($L_{\text{NS}} = 0$) waveguides.

\subsubsection{Nonlinear regime}

The high-incident-power regime ($P_\text{in} \gtrapprox 3~\text{mW}$), on the other hand, is affected by the presence of the nonlinear loss. To estimate the effect of the nonlinear loss on the devices, we calculated the normalized transmittance as
\begin{align}
T_\text{N} = \frac{T_\text{nl}}{T_\text{l}} = \frac{P_\text{out}}{P_\text{out}^{\star}}. \label{eq:TN}
\end{align}
The normalized transmittance represents the ratio of the nonlinear transmittance ($T_\text{nl} = P_\text{out} / P_\text{in}$) to the linear transmittance ($T_\text{l} = P_\text{out}^{\star} / P_\text{in}$), where $P_\text{out}$ is measured by the photodiode and $P_\text{out}^{\star}$ is calculated using Eq.~(\ref{eq:PinPoutNS}). The lower is the normalized transmittance, the higher is the nonlinear loss. $T_\text{N} = 1$ means that only the linear loss is present.

In Section~\ref{sec:results}, we present the results of our measurements, followed by the detailed analysis and interpretation, provided in Section~\ref{sec:discussion}.


\section{Results}
\label{sec:results}

\subsection{Propagation loss}

\begin{figure*}[htbp]
\centerline{\includegraphics[width=15cm]{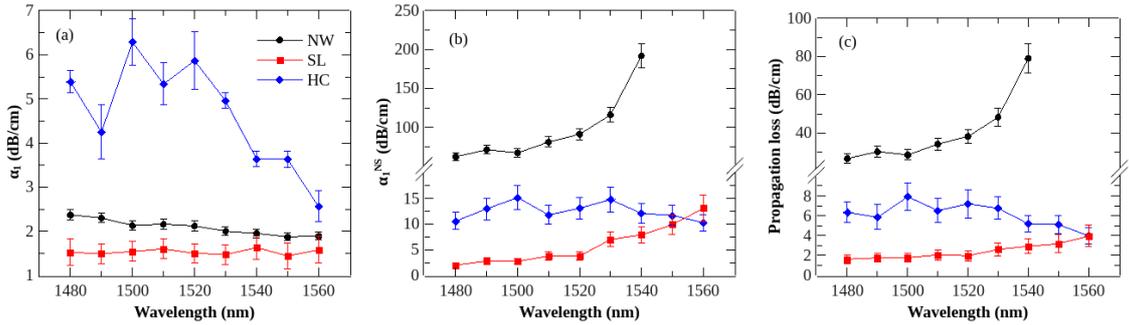}} \caption{\label{fig:alpha1}(color on-line) Linear loss as a function of wavelength: (a) for $2$-$\mu$m-wide waveguides, (b) for $0.6$-$\mu$m-wide nanowire (NW, $\bullet$), $0.9$-$\mu$m-wide strip-loaded (SL, \textbf{\color{red}$\blacksquare$}), and $0.8$-$\mu$m-wide half-core waveguides (HC,  \textbf{\color{blue}$\blacklozenge$}). (c) Total propagation loss for the nanowire, strip-loaded, and half-core waveguides, calculated using $(2 \alpha_1 L_{\text{C}} + \alpha_1^{\text{NS}} L_{\text{NS}})/(2 L_{\text{C}} + L_{\text{NS}})$. The linear loss coefficients are presented in dB/cm in the figure, but the units used in Eqs.~(\ref{eq:PinPout}), (\ref{eq:PinPoutNS}), and~(\ref{eq:eff2}) to (\ref{eq:phi2}) were cm$^{-1}$.}
\end{figure*}

The linear loss coefficients for the $2$-$\mu\text{m}$-wide references waveguides, measured by the Fabry-Perot method, are presented in Fig.~\ref{fig:alpha1}~(a). The propagation losses of this strength are typical for the waveguides of this kind and dimensions. They are caused mainly by scattering from the sidewall imperfections (sidewall roughness)~\cite{porkolab2014low,Dolgaleva2015} or epilayer roughness, as well as wafer defects and, possibly, ion implantation during RIE~\cite{dolgaleva2011compact}. The Fabry-Perot method also allows us to determine the Fresnel reflection coefficient, which was around $R = 0.3$ for all the devices. The linear loss coefficients of the nanosectional parts ($\alpha_1^{\text{NS}}$) are even higher than those of the coupling parts and are presented in Fig.~\ref{fig:alpha1}~(b). The coupling part of the half-core waveguide exhibits the highest loss [see Fig.~\ref{fig:alpha1}~(a)], followed by the nanowire and then strip-loaded, but the nanowire still demonstrates much higher loss in the nanosectional part than that of the half-core and strip-loaded waveguides [Fig.~\ref{fig:alpha1} (b)]. If the overall loss of both the coupling and nanosectional parts is considered, the nanowire exhibits the highest propagation loss, which increases with wavelength, as it is shown in Fig.~\ref{fig:alpha1}~(c). 

The increase of loss with wavelength is expected because longer wavelengths are closer to the cutoff range of the devices~\cite{Dolgaleva2015}. However, for the half-core device, the results on the propagation loss variation with wavelength were not conclusive. It might be possible that a stronger scattering at the input and output facets of the waveguide for shorter wavelengths degraded the internal reflections in the Fabry-Perot cavity, which resulted in an overestimated propagation loss.  Further studies are required to better understand the propagation loss behavior of the half-core geometry.
%

\subsection{Nonlinear loss}

\begin{figure}[h]
\centerline{\includegraphics[width=6cm]{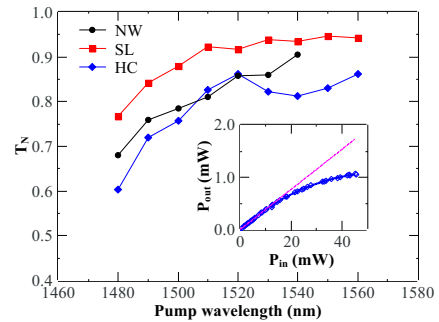}} 
\caption{\label{fig:tn}(color on-line) Nonlinear transmittance as a function of pump wavelength for the TE mode of nanowire (NW, $\bullet$), strip-loaded (SL, \textbf{\color{red}$\blacksquare$}), and half-core (HC, \textbf{\color{blue}$\blacklozenge$}) waveguides. Inset: output power as a function of incident power for the half-core waveguide at the pump wavelength of $1480~\text{nm}$. The solid line is a forth-degree polynomial fitting, and the dashed line is the expected linear behavior if the nonlinear absorption were absent.}
\end{figure}

The output power $P_\text{out}$, measured as a function of the incident power, is represented in Fig.~\ref{fig:tn} (the inset) by the points. The dashed line represents the expected output power $P_\text{out}^{\star}$ in the absence of the nonlinear loss. The nonlinear transmittance spectra of the devices are shown in Fig.~\ref{fig:tn}, where the $T_\text{N}$ values were calculated from Eq.~(\ref{eq:TN}) with $P_\text{in} = 40~\text{mW}$, which is approximately the pump power used in our FWM experiments. One can see that the nonlinear loss increases considerably when the wavelength decreases. The expected mechanisms of the nonlinear loss are the two-photon and three-photon absorption (2PA and 3PA, respectively)~\cite{aitchison1997nonlinear}. The two-photon absorption edge wavelength of the material of the waveguides core, Al$_{0.18}$Ga$_{0.82}$As, is about $1500~\text{nm}$, so, 2PA is expected to occur for $\lambda_{\text{p}} \lessapprox 1500~\text{nm}$, which agrees with Fig.~\ref{fig:tn}. This effect is stronger in the nanowire and half-core waveguides than in the strip-loaded waveguide, which is expected since a smaller modal area would result in higher intensity in the waveguide. At the same time, 2PA and 3PA are intensity-dependent phenomena. Extracting the nonlinear absorption coefficients from $T_\text{N}$ is not straightforward, and one should separate the contributions of the different mechanisms and the different waveguide's parts. This study will be reported in an upcoming publication.
%
\subsection{Nonlinear coefficient and characteristic lengths}

\begin{figure*}[tbp]
\centerline{\includegraphics[width=14cm]{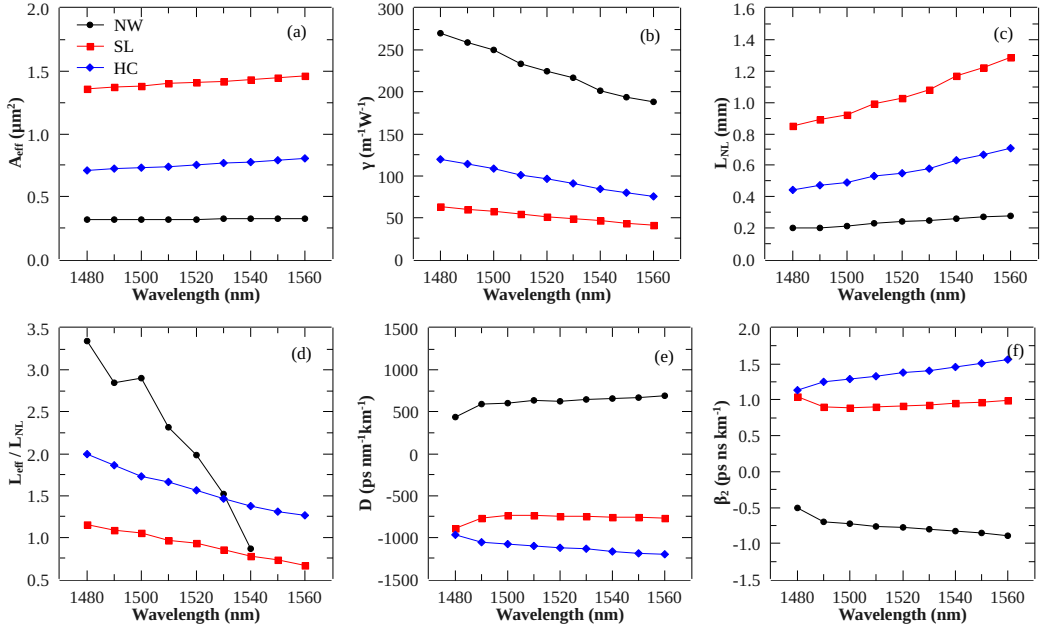}} \caption{\label{fig:characs}(color on-line) (a) Effective TE mode area, (b) nonlinear coefficient, (c) nonlinear length, (d) effective length-nonlinear length ratio, (e) dispersion parameter and (f) GVD coefficient as a function of wavelength for $0.6$-$\mu$m-wide nanowire (NW, $\bullet$), $0.9$-$\mu$m-wide strip-loaded (SL, \textbf{\color{red}$\blacksquare$}), and $0.8$-$\mu$m-wide half-core waveguides (HC, \textbf{\color{blue}$\blacklozenge$}).}
\end{figure*}

An essential characteristic of the device, which indicates the strength of the nonlinear optical interactions in the waveguide, is the nonlinear coefficient. It is given by $\gamma =  2 \pi n_2 / (\lambda_0 A_{\text{eff}})$, where $n_2$ is the nonlinear refractive index and $\lambda_0$ is the incident wavelength. In the wavelength range from $1480~\text{nm}$ to $1560~\text{nm}$, $n_2$ decreases approximately linearly from $2.0 \times 10^{-13}~\text{cm}^2/\text{W}$ to $1.5 \times 10^{-13}~\text{cm}^2/\text{W}$~\cite{aitchison1997nonlinear}. To calculate $\gamma$, we used the values of $A_{\text{eff}}$ obtained from the mode analysis, shown in Fig.~\ref{fig:characs} (a), and the values of $n_2$ from Ref.~\cite{aitchison1997nonlinear}. The results for the nanosectional part of the devices are presented in Fig.~\ref{fig:characs} (b). The device with the highest $\gamma$ is the nanowire, followed by the half-core and then the strip-loaded, as expected based on the different $A_{\text{eff}}$ value for each geometry.
%

Besides the strength of the nonlinear interactions, one also needs to consider the value of the peak power at the waveguide input $P_0$. Both these factors enter the expression for the nonlinear length $L_{\text{NL}} = 1/(\gamma P_0)$, characterizing the effective length of the nonlinear interactions, as presented in Fig.~\ref{fig:characs}~(c). A comparison between the effective length [$L_{\text{eff}} = (1 - e^{- \alpha_1^{\text{NS}} L_\text{NS}})/ \alpha_1^{\text{NS}} ]$  and $L_{\text{NL}}$ is displayed in Fig.~\ref{fig:characs}~(d): the larger the ratio is, the stronger is the nonlinear interaction.

\subsection{Dispersion}

Another factor that influences the FWM process is the dispersion. It mainly affects the phase matching between the pump, signal, and idler beams as they propagate through the waveguide and interact with each other and with the nonlinear medium. One can describe dispersion in terms of the dispersion parameter, given by
\begin{align}
D = - \frac{2 \pi c }{ \lambda^2} \beta_2,\label{eq:D}
\end{align}
where the group velocity dispersion (GVD) coefficient $\beta_2$ is the second derivative of the propagation constant $\beta$ with respect to the frequency $\omega$~\cite{agrawal2000nonlinear}. The overall values of $D$, including the impact of both the material and waveguide dispersions, were obtained from the mode analysis using $\textit{Lumerical Mode Solutions}$ and are shown in Fig.~\ref{fig:characs}~(e). The GVD parameter $\beta_2$ was calculated using Eq.~(\ref{eq:D}) and is presented in Fig.~\ref{fig:characs}~(f). One can conclude from the figure that the nanowire has anomalous dispersion with the absolute value lower than that of the other two geometries. However, none of the devices in this study demonstrated zero GVD in the wavelength range of interest. Previous work has shown that a $0.6$-$\mu$m-wide nanowire with slightly different layer dimensions and material composition exhibits $\beta_2 \approx 0$ for $\lambda \approx 1550~\text{nm}$~\cite{Dolgaleva2015,meier2007group}.

\subsection{Four-wave mixing}

\begin{figure}[htbp]
\centerline{\includegraphics[width=9cm]{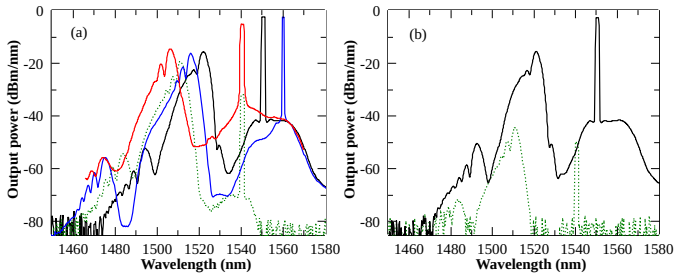}} \caption{\label{fig:tetm}(color on-line) FWM spectra collected for (a) the fundamental TE and (b) TM modes. The dashed green curves are from the $0.6$-$\mu$m-wide nanowire with $L_{\text{NS}}=2~\text{mm}$ when the signal input power was $P_{\text{s}}^{\text{in}}=2.5~\text{dBm}$ and the pump and signal wavelengths were set to $\lambda_\text{p} = 1510~\text{nm}$ and $\lambda_\text{s} = 1540~\text{nm}$, respectively. The solid curves are for the $0.8$-$\mu$m-wide half-core waveguide with $L_{\text{NS}}=1~\text{mm}$, collected at the signal input power $P_{\text{s}}^{\text{in}}=7.5~\text{dBm}$ and the signal and pump wavelengths set to $\lambda_\text{p} = 1505~\text{nm}$ and $\lambda_\text{s} = 1540~\text{nm}$ (red curve), $\lambda_\text{p} = 1515~\text{nm}$ and $\lambda_\text{s} = 1560~\text{nm}$ (blue curve), and $\lambda_\text{p} = 1520~\text{nm}$ and $\lambda_\text{s} = 1550~\text{nm}$ (black curves).}
\end{figure}

To perform a comparative study of the three waveguide geometries, we first verified which guided mode generates the strongest idler output. Fig.~\ref{fig:tetm} presents the result of a set of experiments for the nanowire and half-core devices. For the nanowire (green dashed curve), the fundamental transverse electric (TE) mode [Fig.~\ref{fig:tetm}~(a)] exhibits a higher idler spectral component, generated through FWM, compared to that for the fundamental transverse magnetic (TM) mode [Fig.~\ref{fig:tetm}~(b)]. These results agree with a previous study, which demonstrated that the FWM process in nanowire waveguides is stronger when the polarization of the fundamental mode is TE, as compared to TM~\cite{Dolgaleva2015}. This can be explained by the fact that, for the nanowire, the electric field is more confined in the TE mode than it is in the TM mode [see Figs.~\ref{fig:samples}~(e), and (f)], and by the stronger waveguide dispersion, counterbalancing the material dispersion, characteristic for this polarization. Specifically, zero group velocity dispersion is only achievable for the TE mode for the wavelength range of the FWM experiment. For the half-core waveguide, however, the TE mode is as efficient as the TM mode [black curves in Fig.~\ref{fig:tetm}~(a)~and~(b)]. We decided to set the TE polarization for all the following experiments to have a set of data for all the three geometries at the same polarization. Fig.~\ref{fig:tetm}~(a) also presents FWM spectra of the half-core device for different configurations of pump and signal wavelengths to illustrate their effect on the tunability of the idler wavelength.

The main parameter used for comparing the FWM in various devices is the conversion efficiency, defined as the ratio of the idler power to the signal power, typically expressed in dB. Some works compare the idler power to the signal power, both considered at the output of the waveguide~\cite{choi2018broadband,apiratikul201010,foster2007broad,foster2006broad}. This approach results in the values of the conversion efficiencies modified by the difference in the propagation losses between the signal and idler spectral components. For example, if the idler is generated at the shorter wavelength, the high propagation loss for the signal spectral component will result in lower signal power at the output. The idler-to-signal ratio will appear to be higher. 
An alternative approach, used in other works, is to take the input power of the signal to calculate the idler-to-signal ratio~\cite{yang2018high,wang2018broadband,hu2011ultra,wu2015four,kultavewuti2015low,apiratikul2014enhanced,hansryd2002fiber}. The latter method is used in this study. The conversion efficiency is defined as
\begin{align}
\eta & = 10  \text{log} \left( \frac{P_{\text{i}}^{\text{out}}}{P_{\text{s}}^{\text{in}}} \right), \label{eq:eff1}
\end{align}
where $P_\text{s}^{\text{in}}$ is the signal input power (the power just inside the waveguide, after the coupling, and $P_\text{i}^{\text{out}}$ is the idler power at the output of the waveguide.

The output spectra, collected by the optical spectrum analyzer for the pulsed idler and pump beams, represent a time average, which we represent by $<P(\lambda)>$. If we take into account the repetition rate $f_\text{p}$ of the pump laser and pulse width $\tau$ of the optical pulses, we can calculate the pump or idler temporal peak power from the relationship~\cite{choi2018broadband}
\begin{align}
P_\text{p,i} & = \frac{<P_\text{p,i}>}{\tau f_p},\label{eq:peak}
\end{align}
where $<P_\text{p,i}> = \int_{-\infty}^{+\infty} <P_\text{p,i} (\lambda)> \text{d} \lambda$.

After that, either the peak ($P_\text{i}$) or the average ($<P_\text{i}>$) idler power can be substituted to Eq.~(\ref{eq:eff1}) to calculate the peak ($\eta$) or the average ($<\eta>$) efficiency, respectively. However, the efficiency is usually calculated using the power at the wavelength of maximum amplitude ($\lambda_\text{pk}$), the spectral peak. We, therefore, used the idler spectral peak to calculate the efficiencies in this work, by assuming $<P_\text{i}> = \int_{\lambda_\text{pk} - \delta \lambda}^{\lambda_\text{pk} + \delta \lambda} <P_\text{i} (\lambda)> \text{d} \lambda$ in Eqs.~(\ref{eq:eff1})~and~(\ref{eq:peak}). The integration interval was chosen based on the OSA resolution of $2 \delta \lambda = 1~\text{nm}$. Alternatively, if the integration were performed over all the idler spectrum, the idler average and peak power and the efficiency would have been higher.

\begin{figure}[htbp]
\centerline{\includegraphics[width=9cm]{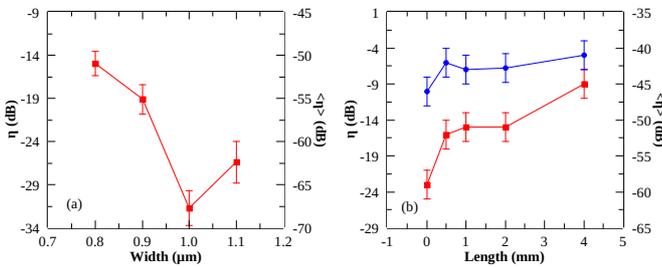}} \caption{\label{fig:lengthwidth}(color on-line) Peak (left axis) and average (right axis) FWM efficiencies for the half-core devices as functions of: (a) nanosectional width with the nanosectional length fixed at $1~\text{mm}$; (b) nanosectional length [see Fig.~\ref{fig:samples} (j)] with the nanosectional width fixed at $0.8~\mu\text{m}$. We set the signal wavelength to $1550~\text{nm}$, and the pump wavelength to $1520~\text{nm}$ (\textbf{\color{red}$\blacksquare$}), and $1535~\text{nm}$ (\textbf{\color{blue}$\bullet$}). The sources of errors, taken into account in calculating the data point errors, are the uncertainty in determining the signal input power and the reproducibility error in coupling the beam into the waveguide and the OSA.}
\end{figure}

Based on the simulation results~\cite{awan2018fabrication}, we identified the parameter space for the waveguide dimensions and fabricated devices with different widths and lengths. After that, preliminary DFWM experiments were performed to determine which dimensions resulted in the highest conversion efficiency for each waveguide geometry. Fig.~\ref{fig:lengthwidth} (a) displays the FWM conversion efficiency of the half-core waveguide with $L_{\text{NS}}=1~\text{mm}$ as a function of the width of the nanosection, indicating that the efficiency is higher for narrower devices. This is expected because the smaller the modal area, the stronger the nonlinear interaction for the same input power levels. Based on this result, the narrowest available device of each geometry was selected for the comparative studies,\textit{ i.e.}  $0.9$-$\mu\text{m}$-wide for the strip-loaded, $0.8$-$\mu\text{m}$-wide for the half-core and $0.6$-$\mu\text{m}$-wide for the nanowire.

Fig.~\ref{fig:lengthwidth}~(b) presents the FWM efficiency for $0.8$-$\mu\text{m}$-wide half-core devices as a function of the nanosectional length $L_{\text{NS}}$, with the fixed total length ($L = 5.87~\text{mm}$). The FWM efficiency exhibits a slight increase with the increase in the length of the nanosection. $L_{\text{NS}} = 0$ represents the FWM generated only by the $2$-$\mu\text{m}$-wide parts and their respective $200$-$\mu\text{m}$-long tapers. A longer $L_{\text{NS}}$ could increase FWM efficiency because the interaction length becomes longer. However, the propagation loss and dispersion would also increase, limiting FWM efficiency. As a compromise between the two contributing factors, we selected intermediate lengths to perform the rest of the experiments: $L_{\text{NS}} = 1~\text{mm}$ for the strip-loaded and half-core devices. For the nanowire geometry, we tested the waveguides with $L_{\text{NS}} = 1~\text{mm}$, and they exhibited fabrication defects. Since we also fabricated waveguides of other lengths on the same sample, we tested the next available length, $L_{\text{NS}} = 2~\text{mm}$, for which we found defect-free working devices. We chose the 2-mm-long nanowires for the FWM comparative study. 
%

%
\begin{figure}[htbp]
\centerline{\includegraphics[width=9cm]{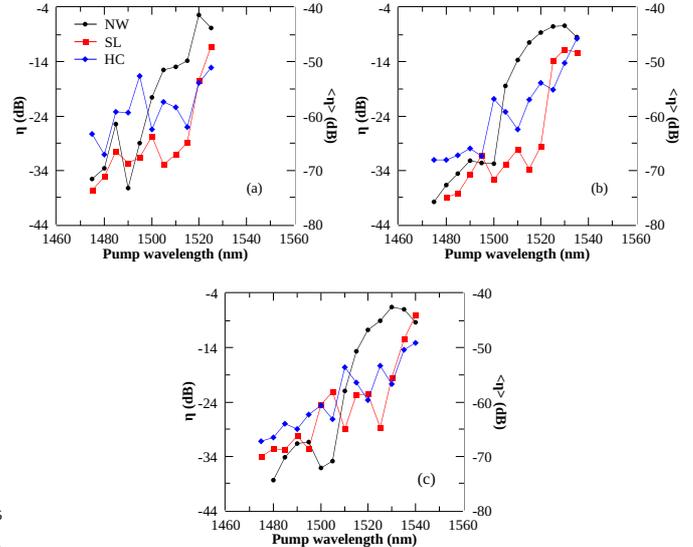}} \caption{\label{fig:effpump}(color on-line) Peak (left axis) and average (right axis) FWM efficiencies as functions of the pump wavelength for the signal wavelength of (a) $1540~\text{nm}$, (b) $1550~\text{nm}$, and (c) $1560~\text{nm}$, for the TE mode in nanowire (NW, $\bullet$), strip-loaded (SL, \textbf{\color{red}$\blacksquare$}), and half-core (HC, \textbf{\color{blue}$\blacklozenge$}) waveguides. The error of each data point (not shown) is $\pm 2~\text{dB}$. The sources of the errors are the same as described in the caption of Fig.~\ref{fig:lengthwidth}.}
\end{figure}

We then performed a series of FWM experiments with different pump wavelengths, ranging from $1475~\text{nm}$ to $1540~\text{nm}$, and with the signal wavelengths of $1540$, $1550$, and $1560~\text{nm}$. The results are presented in Fig.~\ref{fig:effpump}.

\section{Discussion}
\label{sec:discussion}

The maximum efficiencies were $-5~\text{dB}$, $-9~\text{dB}$, and $-8~\text{dB}$, for the nanowire, half-core, and strip-loaded waveguides, respectively, which are among the highest peak efficiencies reported in AlGaAs waveguides. For instance, the achieved values of FWM conversion efficiency, using pulsed pump, were $-10.5~\text{dB}$ for AlGaAs pillar waveguide~\cite{cannon2013all} and $-28~\text{dB}$ for GaAs/AlGaAs waveguide~\cite{apiratikul201010}. For CW-FWM, where both pump and signal were continuous-wave beams, the highest reported conversion efficiencies were $-4~\text{dB}$ for AlGaAsOI micro-ring resonator~\cite{pu2018ultra}, $-6.8~\text{dB}$ for AlGaAs ridge waveguide~\cite{wathen2014efficient}, $-12~\text{dB}$ for AlGaAsOI waveguide~\cite{da2017characterization}, and $-19.8~\text{dB}$ for AlGaAsOS waveguide~\cite{zheng2018high}. The nanowire exhibited superior efficiency at the pump wavelength range above $1500~\text{nm}$, followed by the half-core and then by the strip-loaded waveguides. Below $1500~\text{nm}$, however, the half-core waveguide appears to be the most efficient. The conversion range of these devices, defined as the maximum difference between the signal and idler wavelengths for the observed FWM signals, was $161~\text{nm}$ for the strip-loaded and half-core waveguides and $152~\text{nm}$ for the nanowires, which is $\sim 60~\text{nm}$ broader than the maximum values reported previously for the FWM in AlGaAs waveguides with pulsed pump beam~\cite{Dolgaleva2015,dolgaleva2011compact}.

The linear loss coefficient alone (Fig.~\ref{fig:alpha1}) does not explain the drop in FWM efficiency, observed in Fig.~\ref{fig:effpump} for $\lambda_{\text{p}} \lessapprox 1500~\text{nm}$, because the linear loss increases with the wavelength for the nanowire and strip-loaded, and is approximately constant for the half-core waveguides. 

The nonlinear coefficient could also help to explain why the nanowire has the highest FWM efficiency in Fig.~\ref{fig:effpump}, at least for $\lambda_{\text{p}} > 1500~\text{nm}$. Based on the nonlinear coefficient's inverse wavelength dependence, the nonlinear interaction is expected to be even stronger for $\lambda_{\text{p}} \lessapprox 1500~\text{nm}$. However, this contradicts with the drop in the efficiency of the nanowire observed in our experiments. The reason for this behavior is that the nonlinear absorption also increases for $\lambda_{\text{p}} \lessapprox 1500~\text{nm}$, as we will discuss later.

A simplistic way to account for the influence of the nonlinear coefficient and the input power on the FWM efficiency is to look into the device characteristic lengths. One can conclude from Fig.~\ref{fig:characs} (c) that the lengths of the devices are sufficient to allow for significant nonlinear interactions as the pump and signal propagate through the waveguide (the maximum $L_\text{NL}$ is $1.3~\text{mm}$, whereas $L_{\text{NS}}=1~\text{mm}$ for the strip-loaded and half-core waveguides and $L_{\text{NS}}=2~\text{mm}$ for the nanowire waveguide). This is confirmed by the ratio between $L_{\text{eff}}$ and $L_{\text{NS}}$, presented in Fig.~\ref{fig:characs} (d). One should remark, however, that the higher losses, incurred at the nanowire's nanosectional part at the longer wavelengths, should result in a decrease of the effective length and the FWM efficiency at these wavelengths. However, Fig.~\ref{fig:effpump} shows the opposite. The possible explanation for this is as follows. As the linear loss is caused mainly by scattering, the FWM idler may be generated before a considerable amount of the pump and signal light is scattered. The verification of this can be a subject for further investigations. Besides, the role of the phase matching parameter should also be included in a more sophisticated model for the FWM efficiency, as follows.

A first attempt of comparison between our FWM results and theoretical predictions relies on the following approach. The idler power generated at the output of the waveguide ($P_\text{i}^{\text{out}}$) can be found by solving simplified nonlinear propagation equations for the pump, signal and idler electric fields, as it is detailed in Refs.~\cite{hill1978cw,absil2000wavelength}. Some terms of the resulting expression for $P_\text{i}^{\text{out}}$ correspond to the effect of the phase mismatch, and they can be grouped in a function called phase matching parameter $\phi$ [see Eq.~(\ref{eq:phi})], as it was done in Refs.~\cite{shibata1987phase,tkach1995four}, resulting in $P_\text{i}^{\text{out}} = P_\text{s} (\gamma P_{\text{p}} L_{\text{eff}})^2e^{-\alpha_1 L} \phi$, where $P_\text{s}$ and $P_\text{p}$ are the signal and pump power at the input of the waveguide, respectively. Finally, $P_\text{i}^{\text{out}}$ can be used in Eq.~(\ref{eq:eff1}) to express the FWM efficiency in dB as
\begin{align}
\eta & = 10 \text{log} \left[ \left (\gamma P_{\text{p}} L_{\text{eff}} \right )^2e^{-\alpha_1 L} \phi \right].\label{eq:eff2}
\end{align}

When the phase mismatch is zero, $\phi = 1$. In the general case, however, the phase matching parameter is given by~\cite{shibata1987phase} 
\begin{align}
\phi & = \frac{\alpha_1^2}{\alpha_1^2 + \Delta \beta^2} \left [1 + \frac{ 4  e^{-\alpha_1 L} \text{sin}^2(\Delta \beta L / 2)}{(1-e^{-\alpha_1 L})^2} \right ],\label{eq:phi}
\end{align}
where $\Delta \beta$ is the phase mismatch between the idler, signal and pump. For the pump powers where the nonlinear refraction is non-negligible, the total phase mismatch is given by $\Delta \beta=~\Delta \beta_\text{lin} + 2 \gamma P_\text{p}$, where the first and second terms correspond to the linear and nonlinear contributions to the phase mismatch, respectively~\cite{hu2011ultra,kultavewuti2015low,Dolgaleva2015}. The linear contribution represents the effect of the material and waveguide dispersion and is given by $\Delta \beta_\text{lin} =~2 \beta_\text{p} - \beta_\text{s} - \beta_\text{i}$, where $\beta_\text{p,s or i}$ is the pump, signal or idler propagation constant. The nonlinear contribution represents the effect of the nonlinear refraction on the total phase mismatch, which can either increase or decrease the magnitude of $\Delta \beta$, depending on the signs of $\gamma$ and $\Delta \beta_\text{lin}$.

We performed a comparative analysis between the experimentally measured FWM conversion efficiency and its values predicted by the theoretical model described by Eq.~(\ref{eq:eff2}), using only the parameters of the device's nanosectional part. The direct comparison between the data and the equation did not result in a good agreement. The possible reason for it may be that we are dealing not with the straight waveguides but with the structures that have sections of various widths. To take this into account, we considered the contribution of each part of the structure and modified Eqs.~(\ref{eq:eff2}) and~(\ref{eq:phi}) into
\begin{align}
\eta & = 10 \text{log}\left [ \left (\sum_i \gamma_{i} P_{\text{p} i} L_{\text{eff} i} \right)^2e^{-\overline{\alpha} L} \phi^{\star} \right]\label{eq:eff3}
\end{align}
and
\begin{align}
\phi^{\star} & = \frac{\overline{\alpha}^2}{\overline{\alpha}^2 + \overline{\Delta \beta}^2} \left [1 + \frac{ 4  e^{-\overline{\alpha} L} \text{sin}^2(\overline{\Delta \beta} L / 2)}{(1-e^{-\overline{\alpha} L})^2} \right ],\label{eq:phi2}
\end{align}
respectively, where $\overline{\alpha} = L^{-1} \sum_i \alpha_i L_i$ and $\overline{\Delta \beta} = L^{-1} \sum_i \Delta \beta_i L_i + \sum_i 2 \gamma_{i} P_{\text{p} i}$. The index $i$ represents each of the three parts of the device displayed in Fig.~\ref{fig:samples}~(j), \textit{i.e.}, two parts with $i = \text{C}$ and one with $i = \text{NS}$.

\begin{figure*}[tbp]
\centerline{\includegraphics[width=17cm]{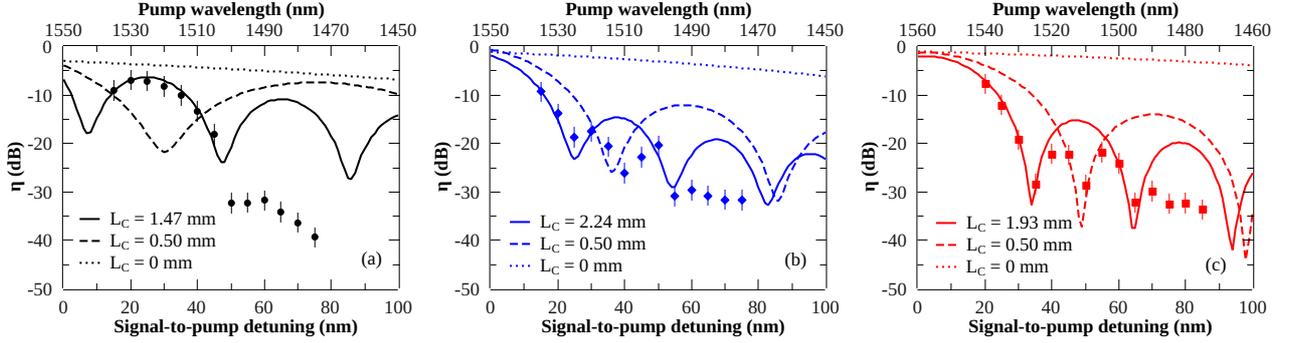}} \caption{\label{fig:sim}(color on-line) Experimental (points) and simulated (solid lines) FWM peak efficiencies for the nanowire (a), half-core (b), and strip-loaded (c) waveguides as a function of the signal-to-pump detuning ($\lambda_{\text{s}} - \lambda_{\text{p}}$) and pump wavelength. For the nanowire and half-core waveguide $\lambda_{\text{s}} = 1550~\text{nm}$, while for the strip-loaded waveguide $\lambda_{\text{s}} = 1560~\text{nm}$. The dashed lines are the results of the simulations performed for different hypothetical values of the coupling lengths $L_{\text{C}}$. Other parameters used in the simulation are: $\alpha_1 = 0.49~\text{cm}^{-1}$, $\alpha_1^{\text{NS}} = 0.90~\text{cm}^{-1}$, $\gamma_{\text{C}} = 60~\text{W}^{-1}\text{m}^{-1}$, and  $\gamma_{\text{NS}} = 180~\text{W}^{-1}\text{m}^{-1}$ for the nanowire waveguide; $\alpha_1 = 0.70~\text{cm}^{-1}$, $\alpha_1^{\text{NS}} = 1.50~\text{cm}^{-1}$, $\gamma_{\text{C}} = 62~\text{W}^{-1}\text{m}^{-1}$, and $\gamma_{\text{NS}} = 90~\text{W}^{-1}\text{m}^{-1}$ for the half-core waveguide; and $\alpha_1 = 0.35~\text{cm}^{-1}$, $\alpha_1^{\text{NS}} = 0.36~\text{cm}^{-1}$, $\gamma_{\text{C}} = 42~\text{W}^{-1}\text{m}^{-1}$, and $\gamma_{\text{NS}} = 51~\text{W}^{-1}\text{m}^{-1}$ for the strip-loaded waveguide.}
\end{figure*}

The FWM efficiencies, calculated using Eqs.~(\ref{eq:eff3}) and~(\ref{eq:phi2}), are plotted in Fig.~\ref{fig:sim} together with the experimental efficiencies. The simulations were performed for the $L_{\text{C}}$ values of our devices and hypothetical shorter coupling lengths. The hypothetical coupling lengths were considered to develop an understanding of how the coupling length affects the conversion efficiency of the FWM process at different values of the signal-to-pump detuning. The results can be explained as follows. If we decrease the length of the coupling part while keeping the length of the nanosectional part the same, the period of the conversion efficiency oscillations as a function of the detuning ($\lambda_{\text{s}} - \lambda_{\text{p}}$) increases. The increase occurs until there are no oscillations observed in the detuning range of interest.
One can thus treat the length of the coupling parts as a design parameter for the FWM in waveguides.

When we use the actual value of $L_{\text{C}}$ in our samples, the theoretical predictions agree fairly well with the experimental results for the longer pump wavelengths, especially for the nanowires. However, for $\lambda_{\text{p}} \lessapprox 1500~\text{nm}$, the experimental points deviate considerably from the theoretically predicted curves. The most significant deviation occurs in the nanowires, indicating the possible impact of the nonlinear loss on the FWM efficiency. Even the linear loss impact on these curves is not completely understood: the $\alpha_1^{\text{NS}}$ values used in Eqs.~(\ref{eq:eff3}) and~(\ref{eq:phi2}) to achieve an agreement between the theory and experiment were lower than the measured ones [shown in Fig.~\ref{fig:alpha1} (b)].

Let us now discuss the nonlinear loss contribution. By comparing Fig.~\ref{fig:tn} with Figs.~\ref{fig:effpump} and \ref{fig:sim}, we conclude that the FWM efficiency drops for $\lambda_\text{p} \lessapprox 1500~\text{nm}$ because of the 2PA increase at those pump wavelengths. The pump photons absorbed by 2PA do not participate in the FWM process. However, the actual spectra of 2PA and 3PA coefficients might not be the same for all three waveguide geometries: the field distribution between the core and cladding layers is different for each geometry. In this work, we only qualitatively assess the nonlinear loss based on the $T_\text{N}$ spectral behavior, and use the 2PA and 3PA results from Ref.~\cite{aitchison1997nonlinear} to support our conclusions. However, calculating the 2PA and 3PA coefficients from $T_\text{N}$ requires further experiments. We are presently performing a separate dedicated study to measure the nonlinear absorption coefficients for the three geometries under different experimental conditions, which we will report in an upcoming publication.

In summary, the half-core is a novel waveguide geometry, and we demonstrated its feasibility by fabricating it in AlGaAs and performing a FWM experiment. This waveguide geometry can be adapted in the situations where the higher FWM efficiency over a broader conversion range is essential. One can implement it in AlGaAs and other material platforms, especially the ones based on III-V semiconductors. The future improvements to the half-core waveguide geometry include the following options. The propagation loss of the coupling part of the half-core waveguide [Fig.~\ref{fig:alpha1}~(a)] was measured to be the highest of all three geometries. However, the propagation loss of the nanosectional part [Fig.~\ref{fig:alpha1}~(b)] and the overall propagation loss [Fig.~\ref{fig:alpha1}~(c)] were much lower than those for the nanowires. A strong scattering present at the input and output facets of the half-core waveguide sample could have been the cause of the higher losses.
Further optimization of the fabrication process should address this issue. Besides, the dispersion is the highest for the half-core geometry [Fig.~\ref{fig:characs}~(e)], and an improvement in the dispersion engineering should increase the FWM efficiency of the half-core waveguide even more. Moreover, there is the possibility of fabricating the half-core waveguide in the form of a microring resonator, which could improve the FWM efficiency by a resonant enhancement~\cite{wu20202d}.

\section{Conclusions}

In conclusion, we proposed and experimentally demonstrated a new AlGaAs waveguide geometry, the half-core waveguide, exhibiting efficient tunable FWM for wavelength conversion with a broad conversion range. We performed a comparative analysis of the new half-core waveguide geometry performances with the better-studied strip-loaded and nanowire waveguides. Compared to the strip-loaded and nanowire waveguides, the half-core is more efficient in the spectral range where 2PA is present. The nanowire waveguide, however, is more efficient in the spectral regions without 2PA. We adapted a theoretical model for calculating FWM efficiency to take into account the various widths and coupling regions in our waveguide structures. The model agrees with the experimental data only in regions without 2PA, which indicates the need to take into account 2PA in modelling FWM conversion efficiency in waveguides.

\section*{Acknowledgment}

The financial support was provided by Canada First Research Excellence Fund award on Transformative Quantum Technologies, the Natural Sciences and Engineering Council of Canada (Discovery and RTI programs) and the Canada Research Chairs program. The epitaxial growth of AlGaAs wafers was supported by a CMC Microsystems award.

\bibliographystyle{model1-num-names.bst}
\bibliography{FWMAlGaAs.bib}

\end{document}